\documentclass{article}
\usepackage{fullpage,amssymb,amsmath,amsthm,algorithm2e}
\usepackage[all]{xy}
\def\eat#1{}
\def\us{\char`\_}
\def\subw#1#2#3{{#1[#2\,..\,#3]}}
\def\abs#1{{|\,#1\,|}}
\def\tree{\mathcal{T}}
\def\lca{\mathrm{lca}}
\newtheorem{theorem}{Theorem}

\newtheorem{lemma}[theorem]{Lemma}
\SetKw{KwFrom}{from} \SetKw{KwContinue}{continue}
\SetKw{KwBreak}{break} \SetKwFunction{KwExtend}{extend}
\SetKwFunction{KwSuffixTree}{make\us suffix\us tree}
\SetKwFunction{KwComputeMP}{compute\us mp}
\SetKwFunction{KwExtendMP}{extend\us mp}
\SetKwFunction{KwComputeRMP}{compute\us rmp}
\SetKwFunction{KwComputeLMP}{compute\us lmp}
\SetKwFunction{KwComputeCMP}{compute\us cmp}

\title{A Minimal Periods Algorithm with Applications}
\author{Zhi Xu}
\date{The University of Western Ontario, \\
Department of Computer Science, \\
Middlesex College, \\
London, Ontario, Canada N6A 5B7 \\
{\tt zhi\us xu@csd.uwo.ca} \\
\medskip
\today}
\begin{document}
\maketitle

\begin{abstract}
Kosaraju in 
``Computation of squares in a string'' briefly described a
linear-time algorithm for computing the minimal squares starting at
each position in a word. Using the same construction of suffix
trees, we generalize his result and describe in detail how to
compute in $O(k\abs{w})$-time the minimal $k$th power, with period
of length larger than $s$, starting at each position in a word $w$
for arbitrary exponent $k\geq2$ and integer $s\geq0$. We provide the
complete proof of correctness of the algorithm, which is somehow not
completely clear in Kosaraju's original paper. The algorithm can be
used as a sub-routine to detect certain types of pseudo-patterns in
words, which is our original intention to study the generalization.
\end{abstract}

\section{Introduction}
A word of the form $ww$ is called a square, which is the simplest
type of repetition. The study on repetitions in words has been
started at least as early as Thue's work \cite{Thue1906} in the
early 1900's. Since then, there are many work in the literature on
finding repetitions (periodicities), which is an important topic in
combinatorics on words. In the early 1980's, Slisenko
\cite{Slisenko1983} described a linear-time algorithm for finding
all syntactically distinct maximal repetitions in a word. Crochemore
\cite{Crochemore1983}, Main and Lorentz \cite{Main&Lorentz1985}
described a linear-time algorithm for testing whether a word
contains a square and thus testing whether a word contains any
repetition. Since a word $w$ of length $n$ may have $\Omega(n^2)$
square factors (for example, let $w={\tt0}^n$), usually only
primitively-rooted or maximal repetitions are computed. Crochemore
\cite{Crochemore1981} described an $O(n\log n)$-time algorithm for
finding all maximal primitively-rooted integer repetitions, where
maximal means that a $k$th power cannot be extend by either
direction to obtain a $(k+1)$th power. The $O(n\log n)$-time is
optimal since a word $w$ of length $n$ may have $\Omega(n\log n)$
primitively-rooted repetitions (for example, let $w$ be a Fibonacci
word). Apostolico and Preparata \cite{Apostolico&Preparata1983}
described an $O(n\log n)$-time algorithm for finding all
right-maximal repetitions, which means a repetition $x^k$ cannot be
extend to the right to obtain a repetition $y^l=x^kz$ such that
$\abs{y}\leq\abs{x}$. Main and Lorentz \cite{Main&Lorentz1984}
described an $O(n\log n)$-time algorithm for finding all maximal
repetitions. Gusfield and Stoye
\cite{Stoye&Gusfield1998,Gusfield&Stoye2004} 
also described several algorithms on finding repetitions. We know
that both the number of distinct squares \cite{Fraenkel&Simpson1998}
and the number of maximal repetitions (also called runs)
\cite{Kolpakov&Kucherov1999} in a words are in $O(n)$. This fact
suggests the existence of linear-time algorithms on repetitions that
are distinct (respectively, maximal). Main \cite{Main1989} described
a linear-time algorithm for finding all leftmost occurrences of
distinct maximal repetitions. Kolpakov and Kucherov
\cite{Kolpakov&Kucherov1999} described a linear-time algorithm for
finding all occurrences of maximal repetitions. For a most-recently
survey on the topic of repetitions in words, see the paper
\cite{Crochemore&Ilie&Rytter2009}.

Instead of considering repetitions from a global point of view,
there are works on a local point of view, which means repetitions at
each positions in a word. Kosaraju in a five-pages extended abstract
\cite{Kosaraju1994} briefly described a linear-time algorithm for
finding the minimal square starting at each position of a given
word. His algorithm is based on an alternation of Weiner's
linear-time algorithm for suffix-tree construction. In the same
flavor, Duval, Kolpakov, Kucherov, Lecroq, and Lefebvre
\cite{Duval&Kolpakov&Kucherov&Lecroq&Lefebvre2004} described a
linear-time algorithm for finding the local periods (squares)
centered at each position of a given word. There may be $\Omega(\log
n)$ primarily-rooted maximal repetitions starting at the same
position (for example, consider the left-most position in Fibonacci
words). So, neither of the two results can be obtained with the same
efficiency by directly applying linear-time algorithms on finding
maximal-repetitions.

In this paper, we generalize Kosaraju's algorithm
\cite{Kosaraju1994} for computing minimal squares. Instead of
squares, we discuss arbitrary $k$th powers and show Kosaraju's
algorithm with proper modification can in fact compute minimal $k$th
powers. Using the same construction of suffix trees, for arbitrary
integers $k\geq2$ and $s\geq0$, we describe in details a
$O(k\abs{w})$-time algorithm for finding the minimal $k$th power,
with period of length larger than $s$, starting at each position of
a given word $w$. ``\emph{The absence of a complete proof prevents
the comprehension of the algorithm} (Kosaraju's algorithm) \emph{in
full details
\ldots}.''\cite{Duval&Kolpakov&Kucherov&Lecroq&Lefebvre2004} In this
paper, we provide a complete proof of correctness of the modified
algorithm. At the end, we show how this $O(k\abs{w})$-time algorithm
can be used as a sub-routine to detect certain types of
pseudo-patterns in words, which is the original intention why we
study this algorithm.

\section{Preliminary}
Let $w=a_1a_2\cdots a_n$ be a word. The \emph{length} $\abs{w}$ of
$w$ is $n$. A \emph{factor} $\subw{w}{p}{q}$ of $w$ is the word
$a_pa_{p+1}\cdots a_q$ if $1\leq p\leq q\leq n$; otherwise
$\subw{w}{p}{q}$ is the \emph{empty word} $\epsilon$. In particular,
$\subw{w}{1}{q}$ and $\subw{w}{p}{n}$ are called \emph{prefix} and
\emph{suffix}, respectively. The \emph{reverse} of $w$ is the word
$w^R=a_n\cdots a_2a_1$. Word $w$ is called a \emph{$k$th power} for
integer $k\geq2$ if $w=x^k$ for some non-empty word $x$, where $k$
is called \emph{exponent} and $x$ is called \emph{period}. The $2$nd
power and the $3$rd power are called \emph{square} and \emph{cube},
respectively.

The \emph{minimal (local) period} $mp_s^k(w)$ larger than $s$ of
word $w$ with respect to exponent $k$ is the smallest integer $m>s$
such that $\subw{w}{1}{km-1}$ is a $k$th power, if there is such
one, or otherwise $+\infty$. For example,
$mp^2_0({\tt0100101001})=3$ and $mp^2_4({\tt0100101001})=5$. The
following results follow naturally by the definition of minimal
period.

\begin{lemma}\label{lemma:mpexpd}
Let $k\geq2$ and $s\geq0$ be two integers and $u$ be a word. If
$mp_s^k(u)\neq+\infty$, then for any word $v$,
  \[mp_s^k(uv)=mp_s^k(u).\]
\end{lemma}
\begin{proof}
Suppose $mp_s^k(uv)<mp_s^k(u)$. We can write $uv=x^ky$ for some
words $x,y$ with $\abs{x}=mp_s^k(uv)>s$. Then $\abs{x^k}=k\cdot
mp_s^k(uv)<k\cdot mp_s^k(u)\leq\abs{u}$ and thus $x^k$ is also a
prefix of $u$. So $mp_s^k(u)\leq\abs{x}=mp_s^k(uv)$, which
contradicts to our hypothesis. So $mp_s^k(uv)\geq mp_s^k(u)$. On the
other hand, if any word $x^k$ is a prefix of $u$, the word $x^k$ is
also a prefix of $uv$. So $mp_s^k(uv)\leq mp_s^k(u)$. Therefore,
$mp_s^k(uv)=mp_s^k(u)$.
\end{proof}

\begin{lemma}\label{lemma:mpshrk}
Let $k\geq2$ and $s\geq0$ be two integers and $u$ be a word. For any
word $v$,
  \[mp_s^k(u)=\begin{cases}mp_s^k(uv),&\textrm{if }\abs{u}\geq k\cdot mp_s^k(uv);\\
  +\infty, &\textrm{otherwise}. \end{cases}\]
\end{lemma}
\begin{proof}
Suppose $mp_s^k(u)\neq+\infty$. By Lemma~\ref{lemma:mpexpd}, it
follows that $mp_s^k(uv)=mp_s^k(u)$ and $\abs{u}\geq k\cdot
mp_s^k(u)=k\cdot mp_s^k(uv)$. So, by contraposition,
$mp_s^k(u)=+\infty$ when $\abs{u}<k\cdot mp_s^k(uv)$. On the other
hand, when $\abs{u}\geq k\cdot mp_s^k(uv)$, we can write $uv=x^kw$
for some words $x,w$ such that $\abs{x}=mp_s^k(uv)$. Then $x^k$ is
also a prefix of $u$ and thus $mp_s^k(u)\neq+\infty$. So, by
Lemma~\ref{lemma:mpexpd}, $mp_s^k(u)=mp_s^k(uv)$.
\end{proof}


The \emph{right minimal period array} of word $w$ with respect to
exponent $k$ and period larger than $s$ is defined by
${}_s^krmp_w[i]=mp_s^k(\subw{w}{i}{n})$ for $1\leq i\leq n$ and the
\emph{left minimal period array} of word $w$ with respect to
exponent $k$ and period larger than $s$ is defined by
${}_s^klmp_w[i]=mp_s^k(\subw{w}{1}{i}^R)$ for $1\leq i\leq n$. For
example,
\begin{align*}
  {}_0^2rmp_{\tt0100101001}&=[3,+\infty,1,2,2,+\infty,+\infty,1,+\infty,+\infty],\textrm{ and} \\
  {}_0^2lmp_{\tt0100101001}&=[+\infty,+\infty,+\infty,1,+\infty,3,2,2,1,5].
\end{align*}

\begin{figure}
\center
  \[\xymatrix{
  & & & +\infty,root\ar[dl]_{\tt0}\ar[dr]^{\tt1} & & \\
  & & +\infty\ar[dl]_{\tt01}\ar[dr]^{\tt1} & & +\infty\ar[dr]^{\tt0}\ar[d]_{\$} & \\
  & 1\ar[dl]_{\tt01001\$}\ar[d]^{\$} & & +\infty\ar[dl]_{\tt0}\ar[d]^{\$} & +\infty,leaf_{10} & +\infty\ar[dl]_{\tt01}\ar[d]^{\tt1001\$} \\
  1,leaf_3 & 1,leaf_8 & +\infty\ar[dl]_{\tt01}\ar[d]^{\tt1001\$} & +\infty,leaf_9 & +\infty\ar[dl]_{\tt01001\$}\ar[d]^{\$} & 2,leaf_5 \\
  & +\infty\ar[dl]_{\tt01001\$}\ar[d]^{\$} & 2,leaf_4 & +\infty,leaf_2 & +\infty,leaf_7 & \\
  3,leaf_1 & +\infty,leaf_6 & & & &
  }\]
\caption{Suffix tree for $\tt0100101001$ with $mp_0^2(\tau(v))$ on
each node $v$}\label{figure:suffixtree}
\end{figure}
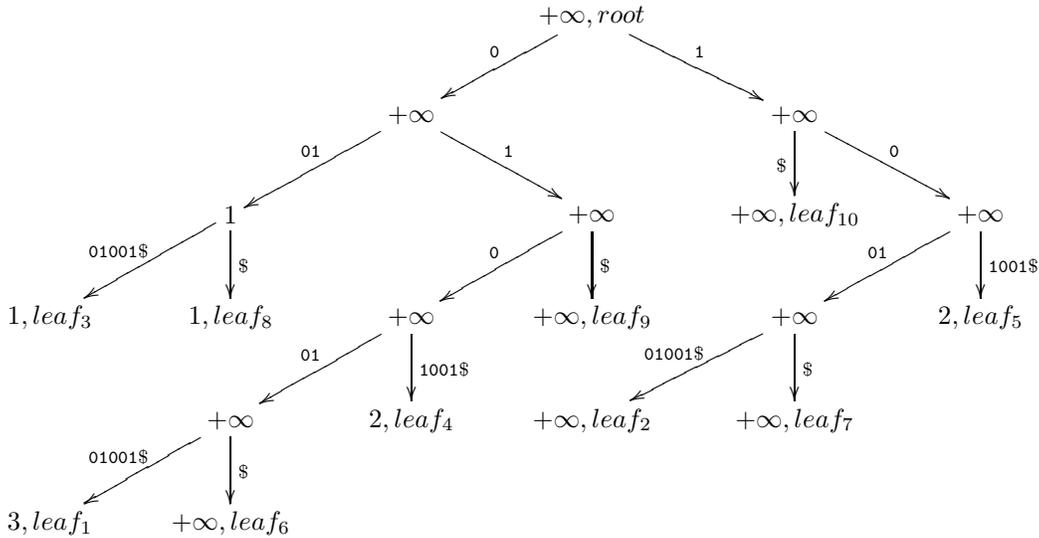

A \emph{suffix tree} $\tree_w$ for a word $w=\subw{w}{1}{n}$ is a
rooted tree with each edge labeled by a non-empty word that
satisfies
\begin{quote}
\begin{enumerate}
  \item each internal node, other than the root, has at least two children,
  \item each label on edge from the same node begins with a different letter, and
  \item there are exactly $n$ leaves $leaf_i$ and
  $\tau(leaf_i)=\subw{w}{i}{n}\cdot\$$ for $1\leq i\leq n$,
\end{enumerate}
\end{quote}
where character $\$$ is a special letter not in the alphabet of $w$
and function $\tau$ is defined at each node $v$ as the concatenation
of the labels on edges along the path from the root to the node $v$.
By definition, a suffix tree for a word $w$ is unique up to renaming
nodes and reordering among children. A suffix tree for the word
$\tt0100101001$ is illustrated in Figure~\ref{figure:suffixtree}.
For more details on suffix tree, see the book
\cite[Chap.~5--9]{Gusfield1997}.

We denote by $p(v)$, or more specifically by $p_{\tree_w}(v)$, the
father of node $v$ in the tree $\tree_w$. Node $x$ is called an
\emph{ancestor} of node $y$ if either $x$ is the father of $y$ or
$x$ is an ancestor of $y$'s father. When node $x$ is an ancestor of
node $y$, node $y$ is called a \emph{descendent} of node $x$. If
node $x$ is a common ancestor of nodes $y$ and $z$ in $\tree_w$, by
the definition of suffix tree, then $\tau(x)$ is a common prefix of
$\tau(y)$ and $\tau(z)$. We denote by $\abs{v}$ the
\emph{node-depth} of node $v$ in $\tree_w$, which is the number of
edges along the path from the root to the node $v$. The node-depth
of the root is $0$ and, for any node $v$, the node-depth $\abs{v}$
is less than or equal to $\abs{\tau(v)}$, which is called the
\emph{depth} of node $v$ in $\tree_w$ and is denoted by $\delta(v)$.
We denote by $\lca(u,v)$ the lowest common ancestor of nodes $u$ and
$v$ in a tree, which is the common ancestor of $u$ and $v$ with the
largest node-depth. After a linear-time preprocessing, the lowest
common ancestor of any pair of nodes in a tree can be found in
constant time \cite{Harel&Tarjan1984,Schieber&Vishkin1988}.


\begin{lemma}\label{lemma:lastedge}
Let $\tree_w$ be the suffix tree of word $w$. If $leaf_i$ and
$leaf_j$ are two leaves such that $i>j$, then the label on the edge
from $p(leaf_i)$ to $leaf_i$ is not longer than the label on the
edge from $p(leaf_j)$ to $leaf_j$.
\end{lemma}
\begin{proof}
Let $n=\abs{w}$ and words $e_i,e_j$ be the labels on the edges from
$p(leaf_i)$ to $leaf_i$ and from $p(leaf_j)$ to $leaf_j$,
respectively. We now prove $\abs{e_i}\leq\abs{e_j}$. Since $i>j$, by
definitions, we can write $\tau(leaf_j)=x\tau(leaf_i)$ for some word
$x$ and thus
  \[\tau(p(leaf_j))e_j=\tau(leaf_j)=x\tau(leaf_i)=x\tau(p(leaf_i))e_i.\]
If $\abs{e_j}\geq\delta(leaf_i)$, then
$\abs{e_i}\leq\delta(leaf_i)\leq\abs{e_j}$. Otherwise, we can write
$\tau(leaf_i)=ye_j$ for some word $y$ and thus $\tau(p(leaf_j))=xy$.
Let $leaf_k$ be another leaf that is a descendent of $p(leaf_j)$.
Then we can write $\tau(leaf_k)=\tau(p(leaf_j))z=xyz$ for some word
$z$ such that $z$ and $e_j$ are different at the first letter. The
word $yz$ is a suffix of $w$ and the longest common prefix of the
two words $\tau(leaf_i)=ye_j$ and $yz$ is $y$. So there is an
ancestor $v$ of $leaf_i$ such that $\tau(v)=y$ and thus
$\delta(p(leaf_i))\geq\abs{y}$. But
$\tau(p(leaf_i))e_i=\tau(leaf_i)=ye_j$. Therefore,
$\abs{e_i}\leq\abs{e_j}$.
\end{proof}

A suffix tree for a given word $w$ can be constructed in linear time
\cite{Weiner1973,McCreight1976,Ukkonen1992}. Both Kosaraju's
algorithm \cite{Kosaraju1994} for computing ${}_0^2rmp_w$ and our
modification on his algorithm for computing ${}_s^krmp_w$ and
$_s^klmp_w$ for arbitrary $k\geq2$ and $s\geq0$ are based on
Weiner's linear-time algorithm \cite{Weiner1973} for constructing
the suffix tree $\tree_w$. So we briefly describe Weiner's algorithm
here.

Weiner's algorithm extends the suffix tree by considering the suffix
$\subw{w}{n}{n}$, \ldots, $\subw{w}{2}{n}$, $\subw{w}{1}{n}$ and
adding $leaf_n$, \ldots, $leaf_2$, $leaf_1$ into the suffix tree
incrementally. After each extension by $\subw{w}{i}{n}$, the new
tree is precisely the suffix tree $\tree_\subw{w}{i}{n}$. The
algorithm is outlined in Algorithm~\ref{figure:weiner}. By using
indicator vectors and inter-node links, the total time to locate
each proper position $y$ at lines~9--10 can be in $O(n)$. Since how
to locate the $y$ is not quite relevant to the algorithm we will
present later, we omit the details here.

\begin{algorithm}
  \SetLine
  \linesnumbered
  \KwIn{a word $w=\subw{w}{1}{n}$.}
  \KwOut{the suffix tree $\tree_{w}$.}
  \Begin(function \KwSuffixTree{$w$}){
    construct $T_n=\tree_{\subw{w}{n}{n}}$ \;
    \For{$i$ \KwFrom $n-1$ \KwTo $1$}{
      \tcp{assert: $T_i=\tree_{\subw{w}{i}{n}}$}
      $T_i\longleftarrow$\KwExtend{$T_{i+1}$, $\subw{w}{i}{n}$} \;
    }
    \Return{$T_1$} \;
  }
  \Begin(function \KwExtend{$tree$, $\subw{word}{i}{n}$}){
  \tcp{we assume $tree=\tree_{\subw{word}{i+1}{n}}$}
    find the proper position $y$ in $tree$ to insert the new node $leaf_i$ \;
    if needed, split an edge $x\to z$ to two $x\to y,y\to z$ by adding a new node $y$ \;
    create and label the edge $y\to leaf_i$ by $\subw{word}{i+\abs{\tau(y)}}{n}\cdot\$$ \;
  }
  \caption{Framework of Weiner's algorithm for constructing suffix tree}
  \label{figure:weiner}
\end{algorithm}

Once a node $v$ is created, although the node-depth $\abs{v}$ may
change in later extensions by splitting on an edge in the path from
the root to node $v$, the depth $\delta(v)$ will never change in
later extensions in a suffix tree. So we assume the depth
$\delta(v)$ is also stored on the node $v$ in the suffix tree and
can be accessed in constant time. The update of $\delta(v)$ only
happens when $v$ is created and can be computed by
$\delta(v)=\delta(p(v))+\abs{u}$, where $u$ is the label on the edge
from $p(v)$ to $v$. So computing and storing the information
$\delta$ will not increase the computational complexity of the
construction of a suffix tree.

\section{The algorithm for computing ${}_s^krmp_w$ and ${}_s^klmp_w$}
First we show that how the minimal period $mp_s^k(w)$ can be
obtained from the suffix tree $\tree_w$ in linear time
$O(\abs{w}/\min\{s,mp_0^k(w)\})$. In particular, if
$s=\Omega(\abs{w})$ and $w$ satisfies $mp_0^k(w)=\Omega(\abs{w})$,
then the algorithm compute $mp_s^k(w)$ in constant time, which is
one of the essential idea in the computing of ${}_s^krmp_w$ and
${}_s^klmp_w$.

\begin{lemma}\label{lemma:mpcret}
Let $k\geq2$ and $s\geq0$ be two integers and $\tree_w$ be the
suffix tree of a word $w$. Then $mp_s^k(w)$ can be computed in
$O\left({\abs{w}}/\min\{s,mp_0^k(w)\}\right)$ time.
\end{lemma}
\begin{proof}
Let $n=\abs{w}$. There is an
$O\left({n}/\min\{s,mp_0^k(w)\}\right)$-time algorithm to compute
$mp_s^k(w)$. First along the path from the $leaf_1$ to the root, we
find the highest ancestor $h$ of $leaf_1$ such that
$\delta(h)\geq(k-1)(s+1)$. Since $\delta(root)=0$, node $h$ always
has a father and $\delta(p(h))<(k-1)(s+1)$. Then we find the least
common ancestor of $leaf_1$ with any other leaf $leaf_i$ that is a
descendent of $h$ and check whether the equation
\begin{equation}\label{equation:mp}
  \delta(\lca(leaf_1,leaf_i))\geq(k-1)(i-1)
\end{equation}
holds. If no $leaf_i$ satisfies (\ref{equation:mp}), then
$mp_s^k(w)=+\infty$; otherwise, $mp_s^k(w)=i-1$, where $i$ is the
smallest $i$ that satisfies (\ref{equation:mp}). The algorithm is
presented in Algorithm~\ref{figure:mp}.

\begin{algorithm}
  \SetLine
  \linesnumbered
  \KwIn{a suffix tree $tree=\tree_{\subw{w}{1}{n}}$ and two integers $s\geq0$, $k\geq2$.}
  \KwOut{the minimal period $mp_s^k(w)$.}
  \Begin(function \KwComputeMP{$tree$, $s$, $k$}){
    \lIf{$k(s+1)>n$}{\Return{$+\infty$}} \lElse{$h\longleftarrow leaf_1$} \;
    \lWhile{$\delta(p(h))\geq(k-1)(s+1)$}{$h\longleftarrow p(h)$ \;}
    $mp\longleftarrow+\infty$ \;
    \tcp{linear-time preprocessing for constant-time finding $\lca$}
    preprocessing the tree rooted at $h$ for $\lca$ \;
    \ForEach{leaf $leaf_i$ being a descendent of $h$ other than $leaf_1$}{
      \If{$\delta(\lca(leaf_1,leaf_i))\geq(k-1)(i-1)$}{
        \tcp{assert: $\subw{w}{1}{i-1}$ is a period of the word $w$}
        \lIf{$mp>i-1$}{$mp\longleftarrow i-1$} \;
      }
    }
    \Return{$mp$} \;
  }
  \caption{Algorithm for computing $mp_s^k(w)$ by using the suffix tree $\tree_w$}
  \label{figure:mp}
\end{algorithm}

Now we prove the correctness of this algorithm. First we observe
that $w=x^ky$ for some non-empty word $x$, if and only if the common
prefix of $\subw{w}{1}{n}$ and $\subw{w}{\abs{x}+1}{n}$ is of length
at least $(k-1)\abs{x}$, which means the leaf $leaf_{\abs{x}+1}$
satisfies (\ref{equation:mp}). Furthermore, $\abs{x}>s$, if and only
if $leaf_{\abs{x}+1}$ satisfies
$\delta(\lca(leaf_1,leaf_{\abs{x}+1}))\geq(k-1)(s+1)$, which means
that $leaf_{\abs{x}+1}$ is a descendent of $h$. (Since $h$ has two
descendents, $h$ is not a leaf and thus $h\neq leaf_{\abs{x}+1}$.)
So each time line~8 is executed, if and only if there is a
corresponding prefix of $w$ that is a $k$th power with period of
length $i-1>s$. The minimal length of such period, if any, is
returned and the correctness is ensured.

Now we discuss the computational complexity of this algorithm. Let
$T_h$ be the sub-tree rooted at $h$ and $l$ be the number of leaves
in $T_h$. By the definition of suffix tree, each internal node has
at least two children in $T_h$ and thus the number of internal nodes
in $T_h$ is less than $l$. Furthermore, the node-depth of any leaf
in $T_h$ is also less than $l$. So the computational time of the
algorithm is linear in $l$. (For details on constant-time algorithm
finding lowest common ancestor with linear-time preprocessor, see
\cite{Harel&Tarjan1984,Schieber&Vishkin1988}.) In order to show the
computation is in $O\left(n/\min\{s,mp_0^k(w)\}\right)$-time, it
remains to see $l=O\left(n/\min\{s,mp_0^k(w)\}\right)$. We prove
$l\leq n/\min\{s+1,mp_0^k(w)\}$ by contradiction. Suppose
$l>n/\min\{s+1,mp_0^k(w)\}$. Since there are $l$ leaves
$i_1,i_2,\ldots,i_l$ with the same ancestor $h$, there are $l$
factors of length $t=(k-1)(s+1)$ such that
  \[\subw{w}{i_1}{i_1+t-1}=\subw{w}{i_2}{i_2+t-1}=\cdots=\subw{w}{i_l}{i_l+t-1}.\]
Since $1\leq i_j\leq n$ for $1\leq j\leq l$, by the pigeon hole
principle, there are two indices, say $i_1$ and $i_2$, such that
$0\leq i_2-i_1\leq n/l<\min\{s+1,mp_0^k(w)\}$. Then the common
prefix of $\subw{w}{i_1}{n}$ and $\subw{w}{i_2}{n}$ is of length at
least $t=(k-1)(s+1)>(k-1)(i_2-i_1)$, which means there is a prefix
of $\subw{w}{i_1}{i_1+t-1}=\subw{w}{i_2}{i_2+t-1}=\subw{w}{1}{t-1}$
that is a $k$th power with period of length $i_2-i_1$. Then
$mp_0^k(w)\leq i_2-i_1<mp_0^k(w)$, a contradiction. So the number of
leaves in $T_h$ is $\leq n/\min\{s+1,mp_0^k(w)\}$ and thus the
algorithm is in $O\left(n/\min\{s,mp_0^k(w)\}\right)$-time.
\end{proof}

For a word $w=\subw{w}{1}{n}$, by definitions, the left minimal
period array and the right minimal period array satisfy the equation
  \[{}_s^klmp_w[i]={}_s^krmp_{w^R}[n+1-i],\textrm{ for }1\leq i\leq n.\]
So the left minimal period array of $w$ can be obtained by computing
the right minimal period array of $w^R$. 
Hence in what follows we only discuss the algorithm for computing
the right minimal period array of $w$; the algorithm for computing
the left minimal period array of $w$ follows immediately.

A \emph{suffix tree with minimal periods} ${}_s^k{\tree_w}$ for a
word $w$ is a suffix tree $\tree_w$ with a function $\pi_s^k$, which
is defined at each node $v$ such that $\pi_s^k(v)=mp_s^k(\tau(v))$.
By definitions, once ${}_s^k\tree_w$ is created for a word
$w=\subw{w}{1}{n}$, the ${}_s^krmp_w$ can be obtained by reading the
value $\pi_s^k$ at each leaf in order as follows:
  \[\subw{{}_s^krmp_w}{1}{n} = [\pi_s^k(leaf_1),\pi_s^k(leaf_2),\ldots,\pi_s^k(leaf_n)].\]
The suffix tree with minimal periods satisfies the following
property.

\begin{lemma}\label{lemma:mpbond}
Let $k\geq2$ and $s\geq0$ be two integers and $w$ be a word. For any
node $v$ in the suffix tree with minimal periods ${}_s^k\tree_w$
such that $\pi_s^k(p(v))=+\infty$, then either $\pi_s^k(v)=+\infty$
or $\pi_s^k(v)$ is between
  \[\frac{\delta(p(v))}{k}<\pi_s^k(v)\leq\frac{\delta(p(v))}{k-1}.\]
\end{lemma}
\begin{proof}
Let $v$ be a node in ${}_s^k\tree_w$ such that
$\pi_s^k(p(v))=+\infty$. Since $\tau(p(v))$ is a prefix of $\tau(v)$
and $\pi_s^k(p(v))=+\infty$, by Lemma~\ref{lemma:mpshrk}, it follows
that
  \[\delta(p(v))=\abs{\tau(p(v))}<k\cdot mp_s^k(\tau(v))=k\cdot\pi_s^k(v).\]
Suppose $\pi_s^k(v)\neq+\infty$. The common prefix of
$\subw{\tau(v)}{1}{\delta(v)}$ and
$\subw{\tau(v)}{\pi_s^k(v)+1}{\delta(v)}$ is of length at least
$(k-1)\pi_s^k(v)$. Then $(k-1)\pi_s^k(v)\leq\delta(p(v))$, since
$p(v)$ is the lowest ancestor of $v$ in ${}_s^k\tree_w$. Therefore,
either $\pi_s^k(v)=+\infty$ or
$\delta(p(v))/k<\pi_s^k(v)\leq\delta(p(v))/(k-1)$.
\end{proof}

In what follows, we will show how to construct the ${}_s^k\tree_w$
for a word $w$ with fixed $k$ in linear time by a modified version
of Kosaraju's algorithm \cite{Kosaraju1994}. Kosaraju's algorithm
constructs only ${}_0^2\tree_w$ but our modification can construct
${}_s^k\tree_w$ for arbitrary $s\geq0$ and $k\geq 2$. Both
algorithms are based on the alternation of Weiner's algorithm
\cite{Weiner1973} for constructing suffix tree $\tree_w$. Our
modified algorithm for computing ${}_s^k\tree_w$ is illustrated in
Algorithm~\ref{figure:rmp}, where the added statements for updating
$\pi_s^k$ are underlined. In addition to the suffix tree
$T_i={}_s^k\tree_{\subw{w}{i}{n}}$, auxiliary suffix tree
$A=\tree_{\subw{w}{p}{q}}$ for some proper indices $p,q$ is used.

The main idea is that we use the classic Weiner's algorithm to
construct the underlying suffix tree $\tree_{\subw{w}{i}{n}}$ step
by step. At each step, at most two nodes are created and we update
the $\pi$ values on those new nodes. One possible new node $y$ is
between two nodes $x,z$ when a split on the edge from $x$ to $z$
happens. Since $\pi_s^k(z)$ is already computed, we update
$\pi_s^k(y)$ directly. The other new node is the new leaf $leaf_i$.
When $\pi_s^k(p(leaf_i))\neq+\infty$, we update $\pi_s^k(leaf_i)$
directly. Otherwise, we compute $\pi_s^k(leaf_i)$ by constructing
auxiliary suffix trees. The na\"ive way is to construct
$\tree_{\subw{w}{i}{n}}$ and then to compute
$\pi_s^k(leaf_i)=mp_s^k(\subw{w}{i}{n})$, both of which run in
$O(\abs{\subw{w}{i}{n}})=O(n)$ time. We instead construct a series
of trees $A=\tree_{\subw{w}{i}{j}}$ for some $j$ in such a way that
$mp_s^k(\subw{w}{i}{n})=mp_s^k(\subw{w}{i}{j})$. In addition, the
total cost of constructing the trees $A$ is in $O(n)$ and each cost
of computing $\pi_s^k(leaf_i)=mp_s^k(\subw{w}{i}{j})$ in each $A$ is
in $O(k)$.

\begin{algorithm}
  \SetLine
  \linesnumbered
  \KwIn{a word $w=\subw{w}{1}{n}$ and two integers $s\geq0$, $k\geq2$.}
  \KwOut{the right minimal period array ${}_s^krmp_{w}$.}
  \Begin(function \KwComputeRMP{$w$, $s$, $k$}){
    construct $T_n$ by constructing $\tree_{\subw{w}{n}{n}}$ \underline{with $\pi(root),\pi(leaf_n)\longleftarrow+\infty$} \;
    \underline{$A\longleftarrow empty$, $j\longleftarrow n$, and $d\longleftarrow 0$} \;
    \For{$i$ \KwFrom $n-1$ \KwTo $1$}{
      find the proper position $y$ in $T_{i+1}$ to insert the new node $leaf_i$ \;
      \If{needed}{
        split an edge $x\to z$ to two $x\to y,y\to z$ by adding a new node $y$ \;
        \underline{\lIf{$\delta(y)\geq k\pi(z)$}{$\pi(y)\longleftarrow\pi(z)$} \lElse{$\pi(y)\longleftarrow+\infty$}} \;
      }
      create and label the edge $y\to leaf_i$ by $\subw{w}{i+\abs{\tau(y)}}{n}\cdot\$$ \;
      \tcp{assert: suffix tree part of $T_i$ is $=\tree_\subw{w}{i}{n}$}
      \underline{\lIf{$j-i+1>{2k}d/{(k-1)}$ or $\delta(y)<d/2$}{$A\longleftarrow empty$}} \;
      \tcp{assert: $A=empty$ or ($A=\tree_{\subw{w}{i}{j}}$ and $d/2\leq\delta(p(leaf_i))\leq2d$)}
      \uIf{$\pi(y)\neq+\infty$}{
        \underline{$\pi(leaf_i)\longleftarrow\pi(y)$} \;
        \underline{\lIf{$A=empty$}{\KwContinue}} \;
        \underline{\lElse{$A\longleftarrow$\KwExtend{$A$, $\subw{w}{i}{j}$}}} \;
      }\Else{
        \uIf{$A=empty$}{
          \underline{$d\longleftarrow\delta(y)$ and $j\longleftarrow i+{(k+1)}d/{(k-1)}-1$} \;
          \underline{$A\longleftarrow$\KwSuffixTree{$\subw{w}{i}{j}$}} \;
        }\Else{
          \underline{$A\longleftarrow$\KwExtend{$A$, $\subw{w}{i}{j}$}} \;
        }
        \underline{$\pi(leaf_i)\longleftarrow$\KwComputeMP{$A$, $\max\{s,\delta(y)/k\}$, $k$}} \;
      }
      \tcp{assert: $\forall v\textrm{ in }T_i:\pi(v)=mp_s^k(\tau(v))$ and thus $T_i={}_s^k\tree_{\subw{w}{i}{n}}$}
      {$rmp[i]\longleftarrow\pi(leaf_i)$} \;
    }
    {$rmp[n]\longleftarrow+\infty$ and \Return{$rmp$}} \;
  }
  \caption{Algorithm for computing ${}_s^krmp_{w}$}
  \label{figure:rmp}
\end{algorithm}

\begin{theorem}\label{theorem:correctness}
Let $k\geq2$ and $s\geq0$ be two integers. Function {\tt compute\us
rmp} in Algorithm~\ref{figure:rmp} correctly computes the right
minimal period array ${}_s^krmp_w$ for the word $w$.
\end{theorem}
\begin{proof}
Since each element ${}_s^krmp_w[i]$ is assigned by the value
$\pi_s^k(leaf_i)$ on the leaves of suffix tree $T_i$ with minimal
periods, the correctness of the algorithm relies on the claim
$T_i={}_s^k\tree_\subw{w}{i}{n}$. The algorithm is based on Weiner's
algorithm and the only change is to update the $\pi_s^k$ values. So
the underlying suffix tree of $T_i$ correctly presents the suffix
tree $\tree_\subw{w}{i}{n}$. The update to $\pi_s^k(v)$ only happens
when the node $v$ is created in some $\tree_\subw{w}{i}{n}$. By
definitions, $\pi_s^k(v)=mp_s^k(\tau(v))$ in any expanded suffix
tree ${}_s^k\tree_\subw{w}{j}{n}$ for $j<i$ is equal to $\pi_s^k(v)$
in the suffix tree ${}_s^k\tree_\subw{w}{i}{n}$ in which $v$ is
created. So in order to prove $T_i={}_s^k\tree_\subw{w}{i}{n}$, it
remains to see that the assignment of $\pi_s^k(v)$ for $v$ is
correct when node $v$ is created.

At the beginning, $\tree_\subw{w}{n}{n}$ is a tree of two nodes, the
root and one leaf $leaf_n$. We have
$\pi_s^k(root)=mp_s^k(\tau(root))=mp_s^k(\epsilon)=+\infty$ and
$\pi_s^k(leaf_n)=mp_s^k(\tau(leaf_n))=mp_s^k(\subw{w}{n}{n})=+\infty$.
So the assignments on line~2 of Algorithm~\ref{figure:rmp} is valid
and $T_n={}_s^k\tree_\subw{w}{n}{n}$.

Suppose it is true that $T_{i+1}={}_s^k\tree_\subw{w}{i+1}{n}$ for
some $i$, $1\leq i\leq n-1$, at the beginning of the execution of
lines~5--25. Then on the next execution within the loop at
lines~5--25, there are at most two nodes being created. One possible
new node is $y$, the father of $leaf_i$, and the other is the
$leaf_i$.

For $\pi(y)$ on line~8: if some split happens on an edge from $x$ to
$z$ by adding a new node $y$ and two new edges from $x$ to $y$, from
$y$ to $z$, respectively, then we have $\tau(z)=\tau(y)u$ for some
$u\neq\epsilon$. By Lemma~\ref{lemma:mpshrk},
$mp_s^k(\tau(y))=mp_s^k(\tau(z))$, if $\abs{\tau(y)}\geq k\cdot
mp_s^k(\tau(z))$; otherwise $mp_s^k(\tau(y))=+\infty$. So the
assignments on line~8 of Algorithm~\ref{figure:rmp} is valid.

For $\pi(leaf_i)$ on line~23: consider the value $\pi_s^k$ on the
new leaf $leaf_i$. Since $y=p(leaf_i)$, we have
$\tau(leaf_i)=\tau(y)v$ for some $v\neq\epsilon$. If
$mp_s^k(\tau(y))\neq+\infty$, by Lemma~\ref{lemma:mpexpd}, it
follows that $mp_s^k(\tau(leaf_i))=mp_s^k(\tau(y))$ and thus the
assignment in line~13 of Algorithm~\ref{figure:rmp} is valid. If
$mp_s^k(\tau(y))=+\infty$, then
$mp_s^k(\tau(leaf_i))=mp_s^k(\subw{w}{i}{n})$ is computed with the
assistant of the auxiliary suffix tree $A=\tree_{\subw{w}{i}{j}}$ by
the function {\tt compute\us mp} in Algorithm~\ref{figure:mp}. Since
$y=p(leaf_i)$, by Lemma~\ref{lemma:mpbond},
$mp_s^k(\tau(leaf_i))>\delta(y)/k$ and thus the arguments in calling
{\tt compute\us mp} is valid. To show the assignment on line~23 of
Algorithm~\ref{figure:rmp} is valid, the only thing remains to prove
is that $mp_s^k(\subw{w}{i}{n})=mp_s^k(\subw{w}{i}{j})$.

First we claim that
$\delta(p_{T_{i}}(leaf_{i}))\leq\delta(p_{T_{i+1}}(leaf_{i+1}))+1$,
where the subscript of $p$ specifies in which tree the parent is
discussed. If $p_{T_{i}}(leaf_{i+1})\neq p_{T_{i+1}}(leaf_{i+1})$,
then there is a split on the edge from $p_{T_{i+1}}(leaf_{i+1})$ to
$leaf_{i+1}$ and leaves $leaf_i,leaf_{i+1}$ has the same father in
$T_i$. So leaves $leaf_{i+1},leaf_{i+2}$ has the same father in
$T_{i+1}$ and thus
  $\delta(p_{T_{i}}(leaf_{i}))
  =\delta(p_{T_{i}}(leaf_{i+1}))
  =\delta(p_{T_{i+1}}(leaf_{i+1}))+1.$
If $p_{T_{i}}(leaf_{i+1})=p_{T_{i+1}}(leaf_{i+1})$, then by
Lemma~\ref{lemma:lastedge}, it follows that
  $\delta(p_{T_{i}}(leaf_{i}))
  \leq\delta(p_{T_{i}}(leaf_{i+1}))+1
  =\delta(p_{T_{i+1}}(leaf_{i+1}))+1.$

Then we claim $\delta(y)\leq j-i+1-2d/(k-1)$ holds right before
line~23, where $y=p(leaf_i)$. Consider the last created suffix tree
$A$, then $A\neq empty$. If $A$ is newly created, then
$\delta(p(leaf_i))=d$ and $i=j+1-(k+1)d/(k-1)$. So
$\delta(p(leaf_i))=j-i+1-2d/(k-1)$. Now we assume $A$ extends from a
previous one. In the procedure of extending $A$, both $j$ and $d$
remain the same, exponent $k$ is a constant, the index $i$ increase
by $1$, and the depth $\delta(p_{T_{i}}(leaf_{i}))$ increases at
most by $1$. So $\delta(p_{T_{i}}(leaf_{i}))\leq j-(i+1)+1-2d/(k-1)$
still holds.

Now we prove $mp_s^k(\subw{w}{i}{n})=mp_s^k(\subw{w}{i}{j})$. If
$mp_s^k(\subw{w}{i}{n})=+\infty$, by Lemma~\ref{lemma:mpshrk}, it
follows that
$mp_s^k(\subw{w}{i}{j})=+\infty=mp_s^k(\subw{w}{i}{n})$. Now we
assume $mp_s^k(\subw{w}{i}{n})\neq+\infty$. By
Lemma~\ref{lemma:mpbond}, it follows that
$mp_s^k(\subw{w}{i}{n})=mp_s^k(\tau(leaf_i))\leq\delta(y)/(k-1)$. In
addition, $j-i+1\leq2kd/(k-1)$ always holds when $A\neq empty$. So
the following holds
  \[k\cdot mp_s^k(\subw{w}{i}{n}) 
  \leq\frac{k}{k-1}\left(j-i+1-\frac{2d}{k-1}\right)
  =(j-i+1)+\frac{1}{k-1}(j-i+1)-\frac{2kd}{(k-1)^2} 
  \leq\abs{\subw{w}{i}{j}},\]
and thus by Lemma~\ref{lemma:mpshrk} again
$mp_s^k(\subw{w}{i}{j})=mp_s^k(\subw{w}{i}{n})$. This finishes the
proof $T_i={}_s^k\tree_{\subw{w}{i}{n}}$.
\end{proof}

\begin{theorem}\label{theorem:complexity}
Let $k\geq2$ and $s\geq0$ be two integers. The time complexity of
computing the right minimal period array ${}_s^krmp_w$ for input
word $w$ in Algorithm~\ref{figure:rmp} is $O(k\abs{w})$ .
\end{theorem}
\begin{proof}
Let $n=\abs{w}$. Each assignment to elements in array $rmp$ at
lines~25,27 of Algorithm~\ref{figure:rmp} can be done in constant
time. So the total time of computing $rmp={}_s^krmp_w$ from the
suffix tree $T_1={}_s^k\tree_w$ with minimal periods is in $O(n)$.

The lines~2,5,7,10 of Algorithm~\ref{figure:rmp} constitute exactly
the Weiner's algorithm for constructing the suffix tree $\tree_w$,
which is in $O(n)$-time.

Most of the underlined statements, except lines~15,19,21,23, in
Algorithm~\ref{figure:rmp} can be done in constant time in a
unit-cost model, where we assume the arithmetic operations,
comparison and assignment of integers with $O(\log n)$-bit can be
done in constant time. The number of executions of lines~5--25 is
$n-1$ and thus the total cost of those underlined statements is in
$O(n)$.

Now we consider the computation of line~23. By
Lemma~\ref{lemma:mpcret}, since $A=\tree_{\subw{w}{i}{j}}$, the cost
of each calling to {\tt compute\us mp} in Algorithm~\ref{figure:mp}
is in time linear in
  \[\frac{\abs{\subw{w}{i}{j}}}{\min\{\max\{s,\delta(y)/k\},mp_0^k(\subw{w}{i}{j})\}}
  \leq\frac{j-i+1}{\min\{\delta(y)/k,mp_0^k(\subw{w}{i}{j})\}}.\]
We already showed in the proof of Theorem~\ref{theorem:correctness}
that $mp_0^k(\subw{w}{i}{j})=mp_0^k(\subw{w}{i}{n})$. By
Lemma~\ref{lemma:mpbond}, $mp_0^k(\subw{w}{i}{n})>\delta(y)/k$. In
addition, $j-i+1\leq2kd/(k-1)$ and $\delta(y)\geq d/2$ always hold
when $A\neq empty$. So we have
  \[\frac{j-i+1}{\min\{\delta(y)/k,mp_0^k(\subw{w}{i}{j})\}}
  \leq\frac{2kd/(k-1)}{\min\{d/2k,d/2k\}}=\frac{4k^2}{k-1}.\]
The number of executions of lines~5--25 is $n-1$ and thus the total
cost on line~23 is $O(kn)$.

Now we consider the computation of lines~15,19,21. Those statements
construct a series of suffix trees $A=\tree_{\subw{w}{i}{j}}$ by
calling to {\tt make\us suffix\us tree} and {\tt extend} in
Algorithm~\ref{figure:weiner}. Each suffix tree is initialized at
line~19, extended at lines~15,21, and destroyed at line~11. Suppose
there are in total $l$ such trees, and suppose, for $1\leq m\leq l$,
they are initialized by $A=\tree_{\subw{w}{i_m}{j_m}}$ with
$d_m=\delta(p_{T_{i_m}}(leaf_{i_m}))$ and destroyed when
$A=\tree_{\subw{w}{i_m'}{j_m}}$ such that
$j_m-(i_m'-1)+1>2kd_m/(k-1)$ or
$\delta(p_{T_{i_m'-1}}(leaf_{i_m'-1}))<d_m/2$ . In addition, when
$A\neq empty$, the inequality $j_m-i+1\leq2kd/(k-1)$ always holds
for $i_m\leq i\leq i_m'$. Since the construction of suffix tree in
Algorithm~\ref{figure:weiner} is in linear time, the total cost on
lines~15,19,21 is in time linear in
  \[\sum_{m=1}^l\abs{\subw{w}{i_m'}{j_m}}
  =\sum_{m=1}^l(j_m-i_m'+1)
  \leq\sum_{m=1}^l\frac{2k}{k-1}d_m.\]
First, we consider those trees $A$ destroyed by the condition
$j_m-(i_m'-1)+1>2kd_m/(k-1)$. Then $j_m-i_m'+1=2kd_m/(k-1)$ and
$j_m=i_m+(k+1)d_m/(k-1)-1$ hold, and thus the decrease of $i$ is
$i_m-i_m'=\left(j_m-(k+1)d_m/(k-1)+1\right)-\left(j_m+1-2kd_m/(k-1)\right)=d_m.$
Hence the total cost in this case is
  \[\sum_{j_m-(i_m'-1)+1>\frac{2k}{k-1}d_m}\frac{2k}{k-1}d_m
  =\frac{2k}{k-1}\sum(i_m-i_m')
  \leq\frac{2k}{k-1}\left((n-1)-1\right)
  =O(n).\]
Second, we consider those trees $A$ destroyed by the condition
$\delta(p_{T_{i_m'-1}}(leaf_{i_m'-1}))<d_m/2$. Then
$\delta(p_{T_{i_m'-1}}(leaf_{i_m'-1}))-\delta(p_{T_{i_m}}(leaf_{i_m}))<-d_m/2$.
In the proof of Theorem~\ref{theorem:correctness}, we showed
$\delta(p_{T_{i}}(leaf_{i}))-\delta(p_{T_{i+1}}(leaf_{i+1}))\leq1$.
Since $\delta(p_{T_1}(leaf_1))\geq0$, it follows that the total cost
in this case is
  \[\sum_{\delta(p(leaf_{i_m'-1}))<\frac{1}{2}d_m}\frac{2k}{k-1}d_m
  <\frac{2k}{k-1}\sum\frac{\delta(p_{T_{i_m}}(leaf_{i_m}))-\delta(p_{T_{i_m'-1}}(leaf_{i_m'-1}))}{2}
  \leq\frac{k}{k-1}(n-1)
  =O(n).\]
The only remaining case is that the suffix tree $A$ is not destroyed
even after the construction of $T_1$. This can be avoided by adding
a special character $\pounds$ not in the alphabet of $w$ at the
beginning of $w$. Then for $i=1$ the father of the $leaf_1$ is the
root and thus $A$ is destroyed by the condition $\delta(y)<d/2$. In
addition, $mp_s^k(\pounds\cdot w)=+\infty$ and thus this
modification do not change the computational complexity of this
algorithm. So, the total cost on lines~15,19,21 is $O(n)$.

Therefore, the total cost of the algorithm is
$O(n)+O(n)+O(n)+O(kn)+O(n)$ and thus is in time $O(kn)$. The
algorithm is in linear time when exponent $k$ is fixed.
\end{proof}

\section{Applications --- detecting special pseudo-powers}
In this section, we discuss how the linear algorithm for computing
${}_s^krmp_w$ and ${}_s^klmp_w$ for fixed exponent $k$ can be
applied to test whether a word $w$ contains a particular type of
repetition, called pseudo-powers.

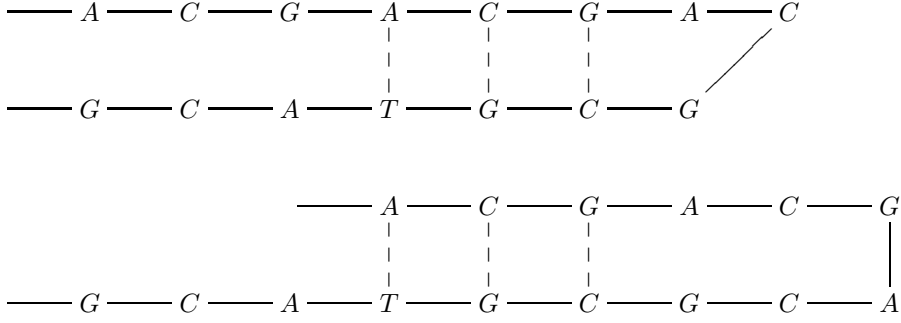
\begin{figure}
\center
  \[\xymatrix{
  \ar@{-}[r] & A\ar@{-}[r] & C\ar@{-}[r] & G\ar@{-}[r] & A\ar@{-}[r] & C\ar@{-}[r] & G\ar@{-}[r] & A\ar@{-}[r] & C\ar@{-}[ld] \\
  \ar@{-}[r] & G\ar@{-}[r] & C\ar@{-}[r] & A\ar@{-}[r] & T\ar@{-}[r]\ar@{--}[u] & G\ar@{-}[r]\ar@{--}[u] & C\ar@{-}[r]\ar@{--}[u] & G \\
  &   &   & \ar@{-}[r] & A\ar@{-}[r]\ar@{--}[d] & C\ar@{-}[r]\ar@{--}[d] & G\ar@{-}[r]\ar@{--}[d] & A\ar@{-}[r] & C\ar@{-}[r] & G \\
  \ar@{-}[r] & G\ar@{-}[r] & C\ar@{-}[r] & A\ar@{-}[r] & T\ar@{-}[r] & G\ar@{-}[r] & C\ar@{-}[r] & G\ar@{-}[r] & C\ar@{-}[r] & A\ar@{-}[u]
  }\]
\caption{An example of hair-pin structure made from a pseudo power
$\tt ACGACGACGCGTACG$ with respect to the Watson-Crick
complementarity}\label{figure:dna}
\end{figure}

Let $\Sigma$ be the alphabet. A function $\phi:\Sigma^*\to\Sigma^*$
is called an \emph{involution} if $\phi(\phi(w))=w$ for all
$w\in\Sigma^*$ and called an \emph{antimorphism} if
$\phi(uv)=\phi(v)\phi(u)$ for all $u,v\in\Sigma^*$. We call $\phi$
an \emph{antimorphic involution} if $\phi$ is both an involution and
an antimorphism. For example, the classic Watson-Crick
complementarity in biology is an antimorphic involution over four
letters $\{\tt A,T,C,G\}$ such that ${\tt A}\mapsto{\tt T}$, ${\tt
T}\mapsto{\tt A}$, ${\tt C}\mapsto{\tt G}$, ${\tt G}\mapsto{\tt C}$.
For integer $k$ and antimorphism $\phi$, we call word $w$ a
\emph{pseudo $k$th power} (with respect to $\phi$) if $w$ can be
written as $w=x_1x_2\cdots x_k$ such that either $x_i=x_j$ or
$x_i=\phi(x_j)$ for $1\leq i,j\leq k$. In particular, we call pseudo
2nd power by \emph{pseudo square} and pseudo 3rd power by
\emph{pseudo cube}. For example, over the four letters $\{\tt
A,T,C,G\}$, word $\tt ACGCGT$ is a pseudo square and $\tt ACGTAC$ is
a pseudo cube with respect to the Watson-Crick complementarity.
Pseudo $k$th power is of particular interest in bio-computing since
a single strand of DNA sequence of this form can itself make a
hair-pin structure as illustrated in Figure~\ref{figure:dna}. A
variation on pseudo $k$th power has also
appeared in tiling problems (see \cite{Beauquier&Nivat1991}). 

Chiniforooshan, Kari and Xu \cite{pseudopower} discussed the problem
of testing whether a word $w$ contains a pseudo $k$th power as a
factor. There is a linear-time algorithm and a quadratic-time
algorithm for testing pseudo squares and pseudo cubes, respectively.
But for general exponent $k$, the known algorithm for testing pseudo
$k$th powers is in $O(\abs{w}^2\log\abs{w})$.

We will show that the particular type of pseudo $k$th powers,
$\phi(x)x^{k-1}$, $x^{k-1}\phi(x)$, and $(x\phi(x))^{\frac{k}{2}}$
(or $(x\phi(x))^{\lfloor\frac{k}{2}\rfloor}x$, if $k$ is odd) can be
tested faster. First we need the following concept. The
\emph{centralized maximal pseudo-palindrome array} ${}^\phi cmp_w$
of word $w$ with respect to an antimorphic involution $\phi$ is
defined by
  \[{}^\phi cmp_w[i] = \max\{m:0\leq m\leq\min\{i,\abs{w}-i\},\phi(\subw{w}{i-m+1}{i})=\subw{w}{i+1}{i+m}\}\textrm{ for }0\leq i\leq\abs{w}.\]
For example, ${}^\phi cmp_{\tt0100101001}=[0,0,0,3,0,0,0,0,2,0,0]$.

\begin{lemma}\label{lemma:cmp}
Let $\phi$ be an antimorphic involution. The centralized maximal
pseudo-palindrome array ${}^\phi cmp_w$ of word $w$ with respect to
$\phi$ can be computed in $O(\abs{w})$ time.
\end{lemma}
\begin{proof}
All maximal palindromes can be found in linear time (for example,
see \cite[pages~197--198]{Gusfield1997}). In exactly the same
manner, by constructing suffix tree $\tree_{w\pounds\phi(w)}$, where
$\pounds$ is a special character not in the alphabet of $w$, the
array ${}^\phi cmp_w$ can be computed in linear time. More
precisely, the algorithm is outlined in Algorithm~\ref{figure:cmp}.

Now we prove the correctness of Algorithm~\ref{figure:cmp}. Let
$n=\abs{w}$ and $\overline{w}=w\pounds\phi(w)$. Then
$\abs{\overline{w}}=2n+1$. By the definition of suffix tree
$\tree_{\overline{w}}$, word $\tau(\lca(leaf_{i+1},leaf_{2n-i+2}))$
is the longest common prefix of
$\tau(leaf_{i+1})=\subw{\overline{w}}{i+1}{2n+1}$ and
$\tau(leaf_{2n-i+2})=\subw{\overline{w}}{2n-i+2}{2n+1}$. Since the
character $\pounds$ does not appear in word $\tau(leaf_{2n-i+2})$
and $\subw{\overline{w}}{1}{i}=\phi(\tau(leaf_{2n-i+2}))$, it
follows that $\tau(\lca(leaf_{i+1},leaf_{2n-i+2}))$ is the longest
word $u$ such that $u$ is a prefix of $\subw{w}{i+1}{n}$ and
$\phi(u)$ is a suffix of $\subw{w}{1}{i}$. (Here $\phi$ is an
antimorphism, so when apply $\phi$, suffix and prefix relations
exchange each other.) This proves the correctness.

Both the construction of suffix tree $\tree_{\overline{w}}$ and the
preprocessing for fast finding $\lca$ is in linear time. In
addition, the computation of $\lca$ for any pair of leaves is
constant after the proprocessing. So the total running time of
Algorithm~\ref{figure:cmp} is in $O(\abs{w})$. \qedhere

\begin{algorithm}
  \SetLine
  \linesnumbered
  \KwIn{a word $w=\subw{w}{1}{n}$ and an antimorphic involution $\phi$.}
  \KwOut{the centralized maximal pseudo-palindrome array ${}^\phi cmp_{w}$.}
  \Begin(function \KwComputeCMP{$w$, $\phi$}){
    $T\longleftarrow$\KwSuffixTree{$w\pounds\phi(w)$} ; \tcp{$\pounds$ is a character not in $w$}
    linear-time preprocessing the tree $T$ for constant-time finding $\lca$ \;
    \For{$i$ \KwFrom $1$ \KwTo $n-1$}{
      $cmp[i]\longleftarrow\delta(\lca(leaf_{i+1},leaf_{2n-i+2}))$ \;
    }
    $cmp[0]\longleftarrow0$ and $cmp[n]\longleftarrow0$ \;
    \Return{cmp} \;
  }
  \caption{Algorithm for computing ${}^\phi cmp_{w}$}\label{figure:cmp}
\end{algorithm}
\end{proof}

\begin{theorem}
Let $k\geq 2$ and $s\geq 0$ be integers and $\phi$ be an antimorphic
involution. Whether a word $w$ contains any factor of the form
$x^{k-1}\phi(x)$ (respectively, $\phi(x)x^{k-1}$) with $\abs{x}>s$
can be tested in $O(k\abs{w})$ time.
\end{theorem}
\begin{proof}
The main idea is first to compute ${}_{\phantom{-1}s}^{k-1}lmp_w$
(respectively, ${}_{\phantom{-1}s}^{k-1}rmp_w$) and ${}^\phi cmp_w$,
and then to compare the two arrays. There is a factor of the form
$x^{k-1}\phi(x)$ (respectively, $\phi(x)x^{k-1}$) with $\abs{x}>s$
if and only if there is an index $i$ such that
${}_{\phantom{-1}s}^{k-1}lmp_w[i]\leq{}^{\phi}cmp[i]$ (respectively,
${}_{\phantom{-1}s}^{k-1}rmp_w[i]\leq{}^{\phi}cmp[i-1]$). More
details of detecting $x^{k-1}\phi(x)$ is given in
Algorithm~\ref{figure:pp1}, and the case of $\phi(x)x^{k-1}$ is
similar.

To see the correctness of Algorithm~\ref{figure:pp1}, we prove that
word $w$ contains any factor of the form $x^{k-1}\phi(x)$ with
$\abs{x}>s$ if and only if
${}_{\phantom{-1}s}^{k-1}lmp_w[i]\leq{}^{\phi}cmp[i]$ holds for some
$i,1\leq i\leq n$, where $n=\abs{w}$. Suppose the inequality
$m={}_{\phantom{-1}s}^{k-1}lmp_w[i]\leq{}^{\phi}cmp[i]$ holds for
some $i$. Then $w$ contains word $\subw{w}{i-(k-1)m+1}{i+m}$ of the
form $x^{k-1}\phi(x)$ as a factor and $\abs{x}>s$. Now suppose $w$
contains a factor $\subw{w}{j}{j+kp-1}$ of the form $x^{k-1}\phi(x)$
for $p=\abs{x}>s$. Then by definitions,
${}_{\phantom{-1}s}^{k-1}lmp_w[j+(k-1)p-1]\leq p$ and
${}^{\phi}cmp[j+(k-1)p-1]\geq p$. So
$m={}_{\phantom{-1}s}^{k-1}lmp_w[i]\leq{}^{\phi}cmp[i]$ holds for
$i=j+(k-1)p-1$.

The computation of ${}_{\phantom{-1}s}^{k-1}lmp_w$ is
$O(k\abs{w})$-time and the computation of ${}^{\phi}cmp$ is
$O(\abs{w})$-time. There are $O(\abs{w})$ comparisons of integers.
So the total running time of Algorithm~\ref{figure:pp1} is in
$O(k\abs{w})$. \qedhere

\begin{algorithm}
  \SetLine
  \linesnumbered
  \KwIn{a word $w=\subw{w}{1}{n}$, an antimorphic involution $\phi$, and two integers $s\geq0$, $k\geq0$.}
  \KwOut{``NO'' if $w$ contains a factor of the form $x^{k-1}\phi(x)$ with $\abs{x}>s$; ``YES'' otherwise.}
  {
    $lmp\longleftarrow$\KwComputeLMP{$w$, $s$, $k-1$} ;
    \tcp{$rmp\longleftarrow$\KwComputeRMP{$w$, $s$, $k-1$} for $\phi(x)x^{k-1}$}
    $cmp\longleftarrow$\KwComputeCMP{$w$, $\phi$} \;
    \For{$i$ \KwFrom $1$ \KwTo $n$}{
      \lIf{$lmp[i]\leq cmp[i]$}{\Return{``NO''} ;
      \tcp{$rmp[i]\leq cmp[i-1]$ for $\phi(x)x^{k-1}$}}
    }
    \Return{``YES''} \;
  }
  \caption{Algorithm for testing whether $w$ contains a factor of the form $x^{k-1}\phi(x)$ with $\abs{x}>s$}
  \label{figure:pp1}
\end{algorithm}
\end{proof}

\begin{theorem}
Let $k\geq 2$ and $s\geq 0$ be integers and $\phi$ be an
antimorphicc involution. Whether a word $w$ contains any factor of
the form $\left(x\phi(x)\right)^{\frac{k}{2}}$ (or
$\left(x\phi(x)\right)^{\lfloor\frac{k}{2}\rfloor}x$ if $k$ is odd)
with $\abs{x}>s$ can be tested in $O(\abs{w}^2/k)$ time.
\end{theorem}
\begin{proof}
The main idea is first to compute ${}^\phi cmp_w$ and then to
enumerate all possible indices and periods. There is a factor of the
specified form as in the theorem if and only if there are $k-1$
consecutive terms greater than $s$ in ${}^\phi cmp_w$ with indices
being arithmetic progression with difference greater than $s$.
The algorithm is given in Algorithm~\ref{figure:pp2}.

To see the correctness of Algorithm~\ref{figure:pp2}, we observe
that $w$ contains a factor of the form
$\subw{w}{i}{j+kp-1}=x\phi(x)x\phi(x)\cdots$ with $p=\abs{x}>s$ if
and only if there are $k$ consecutive terms ${}^\phi
cmp_w[i+p-1],{}^\phi cmp_w[i+2p-1],\ldots,{}^\phi cmp_w[i+(k-1)p-1]$
that are $\geq p>s$.

The computation of ${}^\phi cmp_w$ is $O(\abs{w})$-time and
obviously the remaining part is $O(\abs{w}^2/k)$-time. So the total
running time of Algorithm~\ref{figure:pp2} is in $O(\abs{w}^2/k)$.
\qedhere

\begin{algorithm}
  \SetLine
  \linesnumbered
  \KwIn{a word $w=\subw{w}{1}{n}$, an antimorphic involution $\phi$, and two integers $s\geq0$, $k\geq0$.}
  \KwOut{``NO'' if $w$ contains a factor of the form $\left(x\phi(x)\right)^{\frac{k}{2}}$ (or
$\left(x\phi(x)\right)^{\lfloor\frac{k}{2}\rfloor}x$ if $k$ is odd)
with $\abs{x}>s$; ``YES'' otherwise.}
  {
    $cmp\longleftarrow$\KwComputeCMP{$w$, $\phi$} \;
    \For{$d$ \KwFrom $s+1$ \KwTo $\lfloor n/k\rfloor$}{
      \For{$i$ \KwFrom $0$ \KwTo $d-1$}{
        $consecutive\longleftarrow0$ \;
        \For{$j$ \KwFrom $1$ \KwTo $\lfloor (n-i)/d\rfloor-1$}{
          \lIf{$cmp[i+jd]\geq d$}{$consecutive\longleftarrow consecutive+1$ \;}
          \lElse{$consecutive\longleftarrow0$ \;}
          \lIf{$consecutive\geq k-1$}{\Return{``NO''} \;}
        }
      }
    }
    \Return{``YES''} \;
  }
  \caption{Algorithm for testing whether $w$ contains a factor of the form $\left(x\phi(x)\right)^{\frac{k}{2}}$ with $\abs{x}>s$}
  \label{figure:pp2}
\end{algorithm}
\end{proof}

\section{Conclusion}
We generalized Kosaraju's linear-time algorithm for computing
minimal squares that start at each position in a word, which by our
definition is denoted by the array ${}_0^2rmp_w$. We showed a
modified version of his algorithm that can compute, for arbitrary
integers $k\geq2,s\geq0$, the minimal $k$th powers, with period
larger than $s$, that starts at each position (to the left and to
the right) in a word, which by our definition is denoted by the
right minimal period array ${}_s^krmp_w$ and the left minimal period
array ${}_s^klmp_w$. The algorithm is in $O(k\abs{w})$-time.

The algorithm is based on the frame of Weiner's suffix tree
construction. Although there are other linear-time suffix tree
construction algorithms, such as McCreight's algorithm and Ukkonen's
algorithm, none of the two can be altered to compute minimal period
arrays with the same efficiency, due to the special requirements
that the suffices of the given word are added from the short to the
long and $\pi_s^k(v)$ is only updated when $v$ is created.

We showed the $O(k\abs{w})$-time algorithm for computing minimal
period arrays can be used to test whether a given word $w$ contains
any factor of the form $x^k\phi(x)$ (respectively, $\phi(x)x^k$)
with $\abs{x}>s$. We also discussed an $O(\abs{w}^2/k)$-time
algorithm for testing whether a given word $w$ contains any factor
of the form $\left(x\phi(x)\right)^{\frac{k}{2}}$ (or
$\left(x\phi(x)\right)^{\lfloor\frac{k}{2}\rfloor}x$ if $k$ is odd)
with $\abs{x}>s$. All the word $xx\cdots x\phi(x)$, $\phi(x)x\cdots
xx$, $x\phi(x)x\phi(x)\cdots$ are pseudo-powers. There are
possibilities that some particular type of pseudo-powers other than
the ones we discussed can also be detected faster than the known
$O(\abs{w}^2\log\abs{w})$-time algorithm.


\end{document}